# Range of Normalized Glandular Dose for Mammography Using Patient-Specific Glandular Fractions

Lacey L. Medlock[1], Murtuza S Taqi[1], Bryce Smith[2], Joyoni Dey[1*]

[1]Department of Physics and Astronomy, Medical Physics Program, Louisiana State University, LA
[2]Mary Bird Perkins Cancer Center, Baton Rouge, LA
*deyj@lsu.edu

**Abstract -**

Breast cancer is the most common cancer among women, and mammography remains the primary modality for early detection. Because mammography uses ionizing radiation, accurate estimation of normalized glandular dose (DgN) is important for risk assessment. Recent breast dosimetry models, including TG-282, incorporate population-based glandular tissue distributions; however, patient-specific glandular distributions remain unknown from conventional mammographic projections. In previous work (Smith, Dey et al., 2025), glandular fraction (GF) maps were estimated from a single mammographic projection. While these maps determine glandular path length along each projection ray, they do not uniquely define glandular tissue depth. In this work, we propose a framework for estimating a patient-specific range of DgN from a projection-derived GF map. Using simulated data, glandular tissue was distributed to the top, center, and bottom of the breast volume using Siddon ray-tracing. These configurations preserved the GF map while producing maximum, intermediate, and minimum DgN values. Monte Carlo simulations were performed, and DgN was normalized to entrance air kerma. DgN varied by up to a factor of three solely due to differences in glandular tissue depth, despite identical GF maps and visually indistinguishable projection images. Using randomized realizations derived from TG-282 glandular distributions for Cranio-Caudal (CC) and Medio-Lateral Oblique (MLO) views, dose ratios were calculated relative to central glandular placement. Central placement overestimated DgN by less than 5 and 15 percent on average for MLO and CC distributions respectively, whereas centroid-based placement underestimated dose by up to 25 percent. These results indicate that patient-specific bounds on DgN can be estimated from limited mammographic information and that central placement provides a conservative dose estimate.

## 1. INTRODUCTION

Breast cancer is the most frequently diagnosed cancer among women in the United States and the second leading cause of cancer-related mortality in this population [1]. Prognosis is strongly dependent on stage at diagnosis, therefore, early detection is critical. The United States Preventive Services Task Force recommends biennial screening mammography for women at average risk between 50 and 74 years of age [2, 3]. Given that X-ray mammography is the primary imaging modality for breast cancer screening, an average-risk woman may undergo approximately 13 screening mammograms in her lifetime, contributing to more than 39 million mammographic examinations performed annually in the United States [4].

Although mammography delivers a relatively low radiation dose [5], exposure to ionizing radiation carries a small but non-zero risk of radiation-induced cancer [6]. Consequently, accurate estimation of patient dose during screening remains and important component of risk-benefit analysis in breast cancer screening.

Normalized glandular dose (DgN) estimation methods historically relied on parameterized models that account for anode/filter combinations and assume a homogeneous breast composed of a fixed, typically 50/50 mixture of adipose and fibroglandular tissue [7,8]. However, glandular tissue distribution is spatially heterogeneous and varies substantially, both between patients and within individuals [8-16]. Ignoring this heterogeneity can lead to systemic under- or over-estimation of dose, depending on the true distribution of fibroglandular tissue within the breast.

The breast dosimetry model developed by the 2024 Joint AAPM Task Group 282 / EFOMP Working Group [17] represents a major update to mammography dose estimation by incorporating realistic compressed breast geometries, patient-population based breast density distributions, and separate models for the Cranio-Caudal (CC) and Medio-Lateral Oblique (MLO) views. The model was derived from large patient datasets and Monte Carlo simulations, and was designed to replace earlier dosimetry models that relied on simplified homogeneous breast assumptions.

A key component of the TG-282 work [17] is the fibroglandular tissue distribution model developed by Fedon et al. [18] which used dedicated breast CT images and biomechanical compression simulations to characterize the spatial distribution of glandular tissue in compressed breasts. The study demonstrated that glandular tissue is preferentially concentrated toward the anterior and caudal regions of the breast rather than being uniformly distributed, providing the anatomical basis for the patient-based dosimetry model adopted in TG-282.

In a prior work (Smith, Dey et al [19]), we have shown that a pixel-wise "glandular fraction" map maybe be estimated from a single projection, based on three-tissue modeling of breast. In the current work we propose a patient-specific framework for estimating a *range* of normalized glandular dose (DgN) from the glandular fraction (GF) map, which can be derived from the (single) mammographic projection by our prior method on GF estimation [19]. Since a single mammographic projection does not uniquely determine the three-dimensional distribution of glandular tissue, exact patient-specific dose cannot be recovered from individual standard screening images. However, we can recover an approximate range for each patient from the projection images as well as the derived glandular fraction image.

The goal of this work, then, is not to estimate a single deterministic DgN value, but rather to determine a physically feasible range of DgN values consistent with the estimated glandular fraction map. By identifying the minimum, maximum and intermediate doses under these constraints, we aim to provide clinically meaningful bounds on patient dose that better reflect anatomical uncertainty. We also computed the analytical dose-ratio for randomized trials from the TG-282 MLO and CC with respect to central placement to estimate any resulting biases.

To our knowledge, this is the first study to estimate patient-specific bounds on mammographic DgN directly from glandular fraction maps. The resulting framework provides clinically meaningful dose ranges that explicitly account for uncertainty in glandular tissue depth distribution while remaining compatible with conventional mammographic imaging.

The remainder of this manuscript describes the Monte Carlo simulation framework, estimation of the DgN range variation, reconstruction methodology and validation studies, analytical dose-ratio calculations to assess dose estimation errors due to glandular fraction estimation errors, the comparison of random draw and central placement as well as density-centroid placement, followed by presentation of all the results.

## 2. METHODS

The methods are divided into few parts. In the first part the issue of DgN variation due to fibroglandular fraction distribution was investigated via Monte-Carlo simulations of mammographic projections and dose for different sized breasts for 20% glandular fraction, as well as different glandular fractions (20-50%) for a single breast size. This uses smooth objects generated by TOPAS (Geant4) [20-21]. For each case the fibroglandular tissue is placed at the top, center and bottom and the normalized DgN is calculated via Monte-Carlo. In the subsequent part it is demonstrated that starting with a glandular fraction image, which may be obtained by our prior method from a single projection [19], it is possible to estimate a range of maximum, minimum and an intermediate central DgN by back-projection and concentrating the fibroglandular tissue at the top, bottom and center respectively and then applying TOPAS on these voxelized phantoms. The next third and fourth sections are sections analytically assessing the errors due to errors in estimation of glandular fraction and comparing central placement with random draws from MLO- and CC-view distributions.

### 2.1 Monte Carlo Simulation to Estimate Range of Normalized Glandular Dose (DgN)

Monte Carlo simulations were performed to quantify the range of possible DgN values arising from different depth-wise distributions of fibroglandular tissue within simulated breast objects. Projection images were generated from simulated breast objects in which fibroglandular tissue is concentrated near the top, center, or bottom of the breast.

Dose to the fibroglandular tissue as well as normalized version were calculated.

All simulations were conducted using the TOPAS (Tool for Particle Simulation) framework, a wrapper for the Geant4 Monte Carlo toolkit [20-21]. The simulation pipeline consisted of spectral generation, system geometry modeling, dose scoring within fibroglandular tissue, and normalization by entrance air kerma, each explained below.

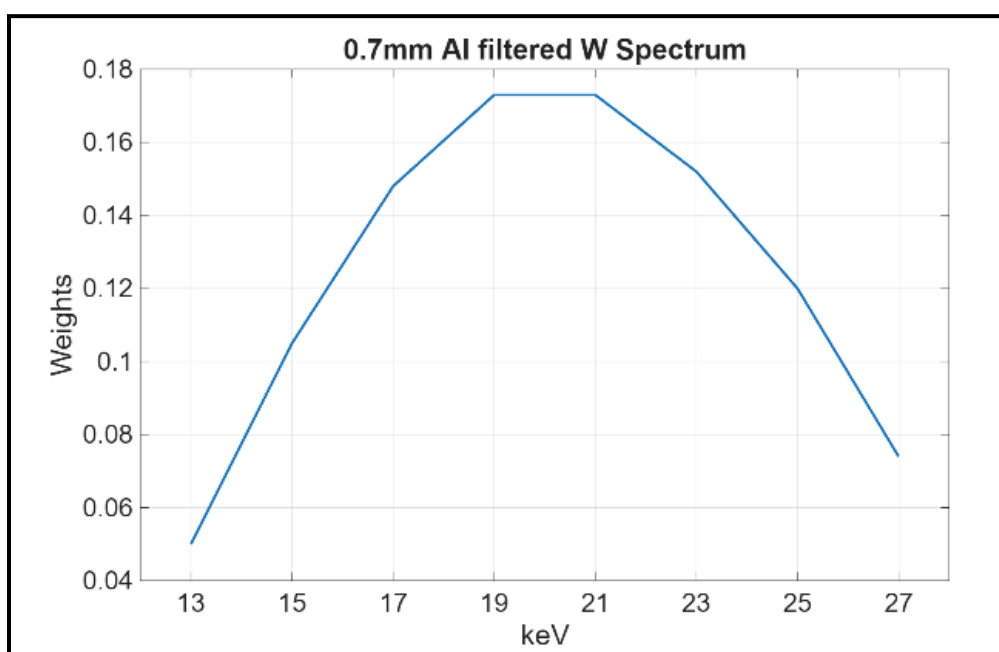

**Figure 1**: Beam weights versus keV for a tungsten anode with 0.7 mm aluminum filter operated at 30kVp. The lowest and highest energy bins are omitted because their contributions to the spectrum account for less than 5%.

***Spectral Generation***

To compute DgN for the various cases, a clinically representative X-ray spectrum was generated. Specifically, a 30-kVp tungsten anode spectrum with 0.7-mm aluminum filtration (W/Al) was modeled, corresponding to a common configuration used in screening mammography. The polychromatic energy spectrum was discretized into 2 keV bins spanning 13-27keV (Figure 1). Energies outside this range were excluded due to their negligible contribution to dose.

***TOPAS geometry and breast models***

TOPAS version 3.7 with the Geant4 Option4 physics list [20-21], which includes the major photon interaction processes relevant to medical X-ray imaging—photoelectric absorption, Compton scatter, and Rayleigh scatter, was used to model a generic mammography system following established implementations in the literature [23-24,19]. The simulated mammography system consisted of an X-ray source, a lead block for beam collimation, a water bath used to simulate body scatter (as commonly employed in mammography simulations [23,19]), Lexan compression paddles positioned above and below the breast, an anti-scatter grid, and a cesium iodide (CsI) detector. The anti-scatter grid was modeled by considering the momentum angle of incident x-ray on the detector and eliminating those incidents above a threshold angle as shown in our previous paper [19]. The threshold angle of 5 deg was chosen as a realistic scatter elimination.

In all simulations, the source-to-detector distance was fixed at 70 cm, consistent with our previous study [19]. Detailed diagrams of the simulated system are provided in [25,19]. This geometry is representative of a clinical mammography system.

The simulated breast was modeled as a semi-elliptical cylinder composed of adipose tissue, fibroglandular tissue, and a 1.45-mm skin layer. Breast diameters ranged from 11.29 to 14.29 cm, with compressed thicknesses from 3.29 to 6.29 cm (Table 1). For all breast diameters, three cases with 20% glandular fraction were modeled. For the 12.29-cm diameter breast, additional simulations were performed for glandular fractions of 30% and 50%. These values were selected to represent the distribution of breast compositions commonly encountered in screening mammography. For each configuration, the fibroglandular tissue was arranged in three depth distributions: concentrated near the top (source side), center, or bottom (detector side) of the breast (Figure 2).

**Table 1. Dimensions of the simulated breast for Monte Carlo modeling in TOPAS**

| Diameter (including skin) | Thickness (including skin) | Chest wall to nipple distance (including skin) |
|---|---|---|
| 11.29 cm | 3.29 cm | 5.645 cm |
| 12.29 cm | 4.29 cm | 6.145 cm |
| 13.29 cm | 5.29 cm | 6.645 cm |
| 14.29 cm | 6.29 cm | 7.145 |

For each 2keV bin, 120 million photons were simulated, providing sufficient signal-to-noise ratio while maintaining feasible computation times. The resultant projections were weighted by the spectrum and added to obtain resultant projections (example shown in Figure 3).

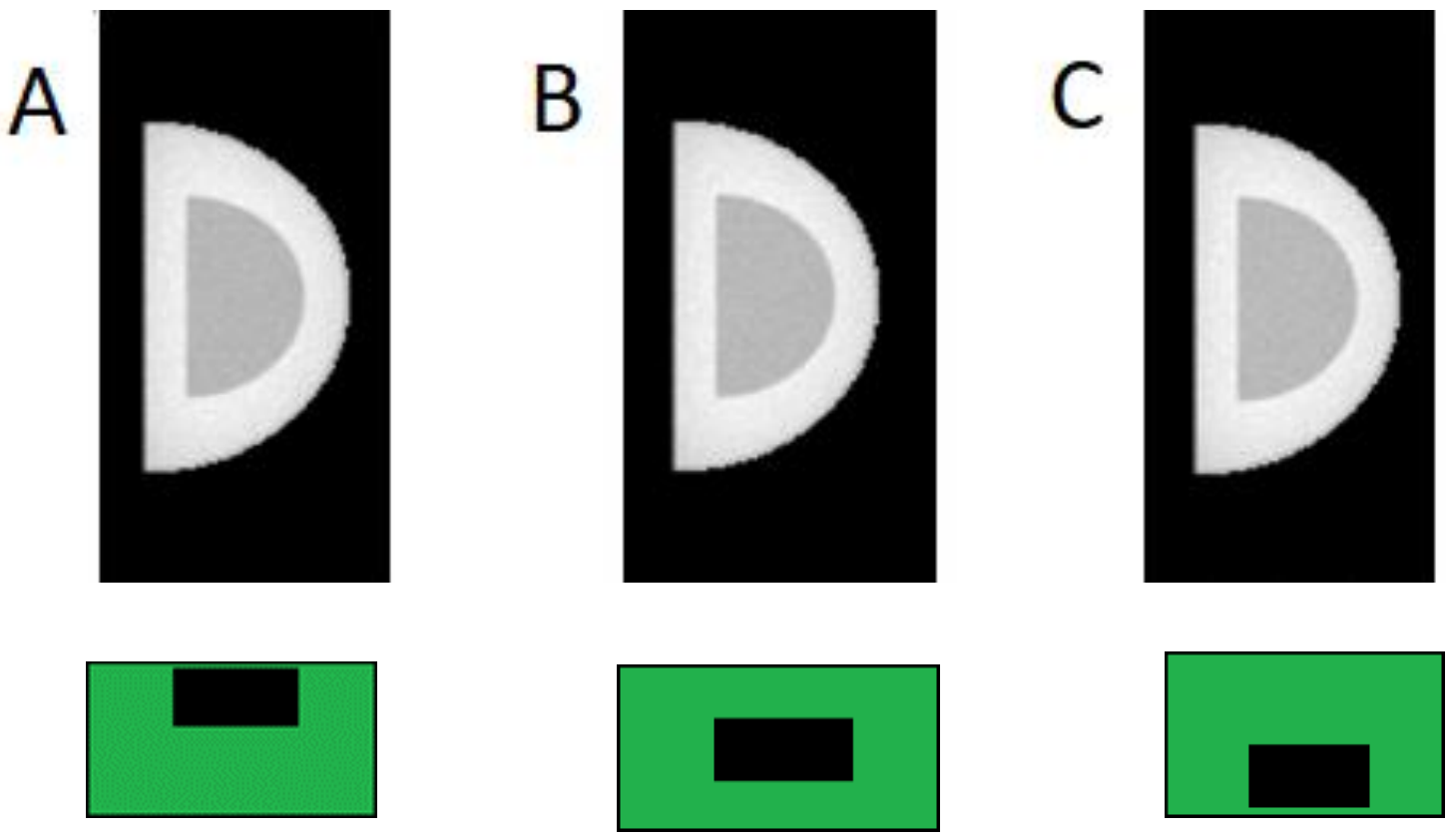

**Figure 2.** An example of projection images for a simulated breast object with fibroglandular tissue concentrated at different depths, as is illustrated by the green and black schematic at the bottom of the image. Fibroglandular tissue concentrated (a) at the top, closest to the X-ray source, (b) in the center of the simulated breast object, (c) at the bottom of the simulated breast object, farthest away from the X-ray source. Despite their visual similarity, the corresponding DgN values differ substantially, illustrating the need for a range-based dose characterization.

Dose (D) to the fibroglandular tissue was recorded, and entrance air kerma was estimated by tallying photons passing through a 3 cm × 3 cm scoring plate positioned on the upper surface of the compression paddle adjacent to the chest wall. Although scoring-plate sizes reported in the literature range from 0.8 cm × 0.8 cm to 1.8 cm × 3 cm [23-24], a larger 3 cm × 3 cm plate was used in this study to better capture beam-intensity variation.

The DgN is calculated as described below.

***DgN calculation***

Entrance surface air kerma was calculated for each energy according to the following equation.

$$K_E = 1.602 \times 10^{-10} \times E \times 10^{-3} \times \frac{\mu_{en}}{\rho}_{air} \times \frac{N}{9cm^2} \qquad (1)$$

where $K_E$ is entrance surface air kerma, E is energy in keV, $\frac{\mu_{en}}{\rho}_{air}$ is the mass-energy absorption coefficient for dry air at energy E, and N is the number of photons that pass through the 3 cm × 3 cm scoring plane. The air kerma for each energy was weighted by the beam weight, as was the dose for each energy. The DgN was then calculated by

$$DgN = \frac{\sum_E D_E W_E}{\sum_E K_E W_E} \quad (2)$$

where $W_E$ is the beam weight for each energy and $D_E$ is the dose for each energy. The sums were performed across all energy bins.

The clinical metrics Mean Glandular Dose (MGD) depends on variety of system specific factors such imaging protocols such as exposure, spectrum. In our study we computed the glandular dose normalized by the entrance Kerma for each case. The clinical metric, MGD can be approximately calculated as $MGD = DgN \times K_E$.

The actual dose is calculated using TOPAS Monte Carlo simulations, with the scoring volume defined as the glandular tissue regions. These regions may be represented either by a homogeneous analytical object or by a labeled, voxelized phantom. Within the scoring volume, TOPAS tracks all particle interactions and accumulates the deposited energy. The mean glandular dose is then calculated by dividing the total deposited energy over the total mass of the glandular tissue.
The TOPAS Monte Carlo setup was used to assess the dose variation associated with glandular tissue placement in regular semi-elliptical cylindrical objects (results in Section 3.1).

In the following section, we describe the generation of breast volumes from glandular fraction images using Siddon ray tracing and the subsequent use of TOPAS for glandular dose calculations on these reconstructed voxelized phantoms.

### 2.2 Dosimetry from Breast Volumes Reconstructed Consistent with Pixel-wise Glandular Fraction Data

To estimate the minimum and maximum possible DgN values corresponding to the glandular fraction image (which can be derived based on [19]), a Siddon ray-tracing algorithm [26,27] was used to back-project a ray from each detector pixel into the breast volume towards the source. Figure 3 provides a simplified, two-dimensional illustration of the reconstruction procedure. The object is positioned between the X-ray source and the detector. A single detector pixel in the projection image is selected as the pixel of interest, and the corresponding ray is back-projected from the detector toward the source. All voxels intersected by this ray are identified and are subsequently assigned as adipose tissue, fibroglandular tissue, or skin, consistent with the glandular fraction image.

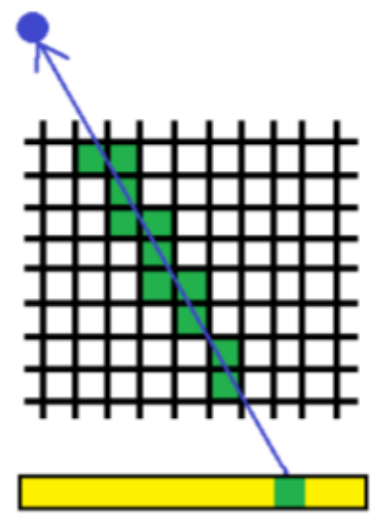

**Figure 3.** Illustration of the Siddon ray-tracing algorithm to distribute the glandular tissue. The yellow bar represents the projection image, with the green box on the yellow bar marking a single pixel of interest. The blue arrow indicates the pixel-to-source backprojection path utilized by the ray-tracing algorithm. Voxels intersected by the ray corresponding to the selected projection pixel are identified and highlighted in darker green.

To isolate the reconstruction code from possible errors in the glandular fraction estimation code [19], a "perfect" glandular fraction map was assumed, specifying the fraction of fibroglandular tissue along each ray. In a later section we analytically assess the dosimetry error due to glandular fraction estimation error.

The glandular fraction map provides the total fibroglandular path length that must be satisfied for each ray, but does not specify how that fibroglandular tissue is distributed in depth. Using this information, voxels intersected by each ray were assigned as fibroglandular tissue or adipose tissue in a manner that either minimized or maximized dose, while preserving the projection image and GF map.

For each projection pixel, the ray-tracing algorithm identifies all voxels intersected by the corresponding ray. The total fibroglandular thickness associated with that ray is calculated as the product of the pixel's GF value and the compressed breast path-length traversed by each ray. Initially, all intersected voxels are assigned as adipose tissue. Voxels are then iteratively assigned as fibroglandular tissue until the cumulative fibroglandular path length along the ray matches the glandular tissue thickness specified by the glandular-fraction map.

Appendix A shows an analytical proof of dose at the glandular tissue being maximized if the tissue is concentrated at the top (entrance) surface of the breast versus distributed along the depth. The minimization when the glandular tissue is at the bottom follows as a corollary.

To generate the maximum dose configuration, fibroglandular tissue is preferentially assigned to voxels closest to the X-ray source along each ray. Beginning at the entrance surface of the breast, voxels traversed by Siddon-ray-tracing are converted from adipose to fibroglandular tissue one at a time until the required glandular path length is achieved. The reconstruction is performed in MATLAB using 1 $mm^3$ voxel size. Conversely, to generate the *minimum* dose configuration, fibroglandular tissue is assigned starting at the deepest voxels along the ray, farthest from the source, and proceeding toward the entrance surface until the same glandular path length constraint is satisfied. In both cases, this procedure assures the reconstructed volume reproduces the original projection image and GF map, while dose distributions are extremized. Similarly for the *central* configuration, adipose voxels were converted to fibroglandular starting from the central line and proceeding up and down until the glandular path length constraint is satisfied.

Once fibroglandular and adipose assignments are completed for all rays, a skin layer is added to the surface of the reconstructed breast volume. The skin is modeled as a uniform 1.45 mm thick layer across the entire surface. However, due to the 1 mm voxel resolution, the outermost voxel layer is assigned as skin, resulting in an effective skin thickness of 1 mm.

Cross sections of representative reconstructed breast volumes are shown in Figure 4. Figure 4a shows the projection image generated in TOPAS for a 12.29 × 4.29 cm breast. Figures 4b and 4c show cross sections through the reconstructed volumes optimized to maximize and minimize dose, respectively. In both cases, the cross section corresponds to the x = 95 mm plane, which bisects the reconstructed object.

In both reconstructed volumes, the proximal (left) boundary of the reconstructed breast object is parallel to the z-axis, corresponding to the chest wall. However, the distal (right) boundary is slightly slanted, as expected from the divergent ray geometry used in the back-projection, arising from the finite source-to-detector distance of 70 cm.

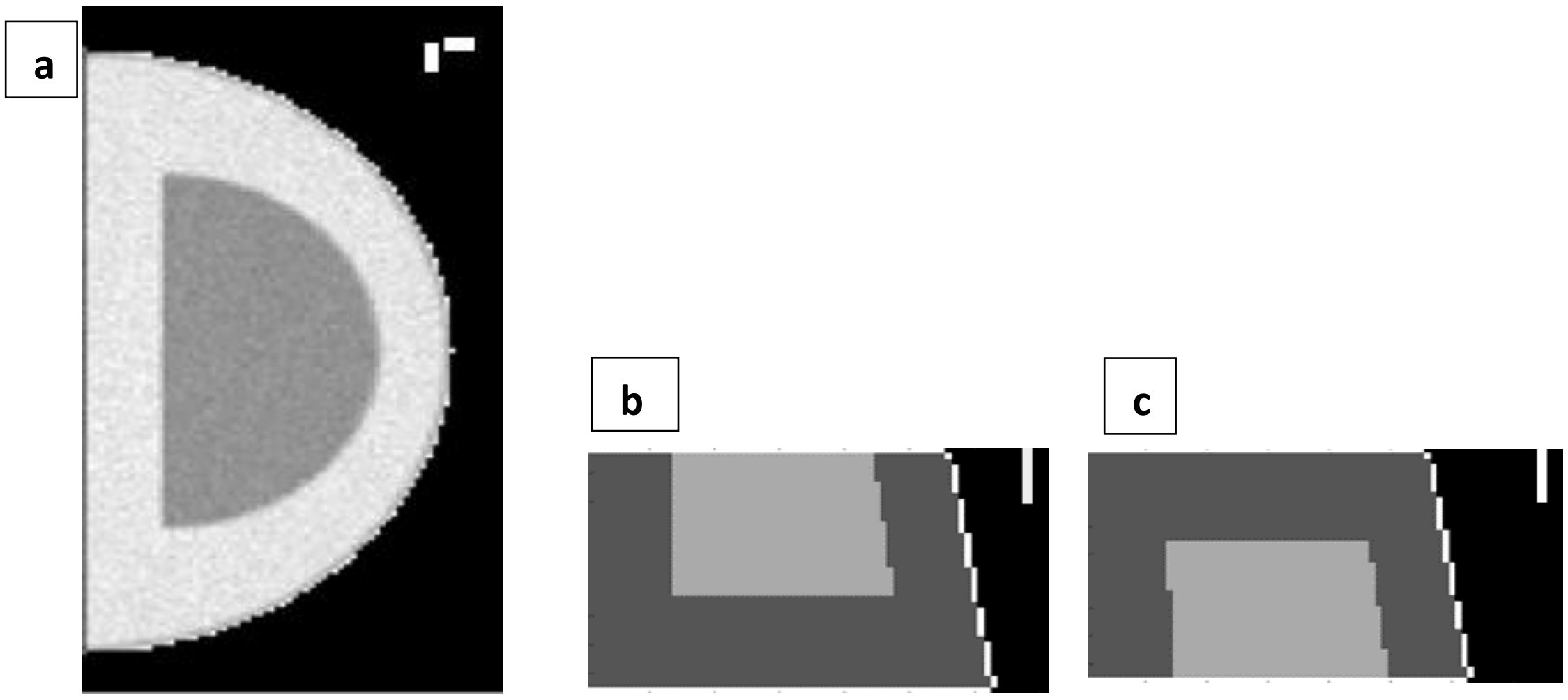

**Figure 4.** Reconstructions for a 12.29 cm (horizontal radius) × 4.29 cm (thick) breast object (a) A projection image generated in TOPAS, with fibroglandular tissue concentrated at the top of the breast. (b) Cross section of the reconstructed volume optimized to maximize dose, shown in the x=95 mm plane. (c) Cross section of the reconstructed volume optimized to minimize dose, shown in the same plane. For all three images, the bars to the right indicates 1cm length.

The dosimetry results for Siddon ray-traced breast models of varying thicknesses are presented in Section 3.2. These results are compared with those from Section 3.1, which were obtained using smooth, analytically defined TOPAS (Geant4-based) generated objects. This comparison quantifies errors introduced by voxelization as well as any additional errors associated with the reconstruction procedure.

**2.3 Analytical Estimation of Dose Error Due to Glandular Fraction Errors for the Central-Placement Case**

The simulations presented here assumed ideal pixel-wise glandular fraction estimation. However, our current estimation methods have an average absolute error of approximately 4.5–4.7%, depending on whether an anti-scatter grid is included or the scatter is removed by software [19]. To evaluate the impact of this uncertainty on dosimetry, we analytically computed the dose ratios for the central-placement case for ±5% glandular fraction errors for breast thicknesses ranging from 3–6 cm and glandular fractions of 20%, 30%, and 50%.

The analytical glandular dose calculation is described next. Since our estimation is pixel-wise, we consider the dose along a single vertical ray of thickness T, with values ranging from 30–60 mm. A voxel size of 0.1 mm was used to adequately capture glandular voxel differences associated with ±5% changes in glandular fraction. The absorbed dose at each voxel is approximately given by (Eqn A.1),

$$D \propto \frac{\frac{\mu_{en}}{\rho} e^{-atten}}{(l+s)^2} \qquad (3)$$

where $atten = \sum \mu_i \Delta l_i$ to account for the linear-attenuation of flux and and $\mu_{en}$ accounts for the energy absorption for adipose and glandular tissue depending on the type of the i-th voxel. The summation extends from the top of the breast to the current voxel. The $l$ denotes the distance from the top of the breast to current voxel and s is the distance from the X-ray source to the top of the breast. The dose deposition at the *glandular tissue-voxels* is then summed up.

The process was repeated for monochromatic spectra from 13–27 keV in 2 keV intervals. The resulting doses were weighted using the spectral distribution shown in Figure 3. and summed to obtain the final dose estimate. Dose ratios corresponding to the nominal glandular fraction and ±5% glandular fraction variations were then computed. The resulting dose ratios for breast thicknesses ranging from 30–60mm is summarized in the Results Section 3.3

### 2.4 Analytical Dose Comparisons between Randomly Distributed and Centrally-placed Glandular tissue cases

Using the approximate analytical dose model described in Eqn. 3 in the previous section, we calculated the dose ratios between randomly distributed glandular tissue cases and the corresponding central-placement case for breast thicknesses of 3–6 cm and glandular fractions of 20%, 30%, and 50%.

Fibroglandular distributions obtained from breast CT data and subsequently compressed using a finite-element model were reported by Fedon et al. [18], and have been included the 2024 AAPM task force report [17]. We adapted the axial distributions for both the CC and MLO views, with the caveat that the normalized density profiles were assumed to approximately retain the same distribution shape for breast thicknesses between 30–60 mm. Since the reported distributions extend to a maximum thickness of 70 mm, the depth axis within the breast was normalized by T/70, where T is the breast thickness in mm. The resulting CC and MLO glandular density distributions as a function of thickness normalized breast depth are shown in Figure 4(a).

Similar to previous section, a vertical ray with pixel size 0.1 mm was considered of length given by the thickness considered and glandular tissue pixel distributions were randomly generated 100 times drawn from the corresponding probability distribution. An example realization for the 60 mm MLO distribution, along with the corresponding central-placement case, is shown in Figure 4(b). The polychromatic glandular dose was then calculated as described in the previous section for each of the 100 cases.

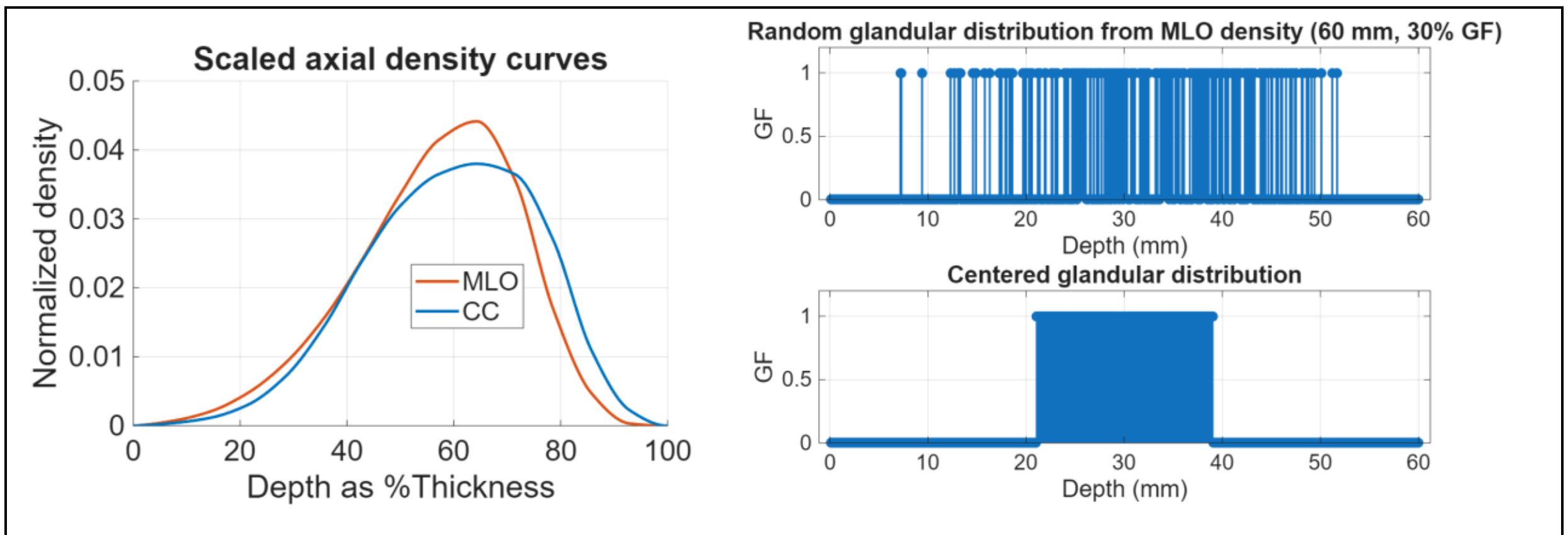

**Figure 5.** (a) MLO and CC Density functions adapted from [17], scaled to percent thickness (b) Example random draw (top) with depth for 60mm thickness, 30% Glandular Fraction. Each marked pixel is assumed to be fully glandular. The bottom shows all the glandular tissue concentrated in the central region.

We also repeated the comparison of the random-placements with the placement at the centroid of the distributions for the MLO and CC cases.

The average (and standard deviation) of dose ratios of the randomly distributed cases (for both MLO and CC views) relative to the central-placement as well as the centroid-placement cases were computed for breast thicknesses of 30–60 mm and glandular fractions of 20%, 30%, and 50%, and are reported in the Results Section.

## 3. RESULTS

### 3.1 Estimation of DgN

Figure 6 summarizes DgN values for breasts of varying thickness at 20% glandularity. There was expected difference between the lower, central and upper placement due to the relative attenuation by the breast tissues and inverse-square law. For all fibroglandular distributions, the DgN decreases as the thickness of tissue increases due to increased attenuation and greater average depth of fibroglandular tissue. However, the ratio between maximum and minimum DgN increases with thickness, from approximately 1.8 for a 3.29 cm thick breast to nearly 3.0 for a 6 cm thick breast.

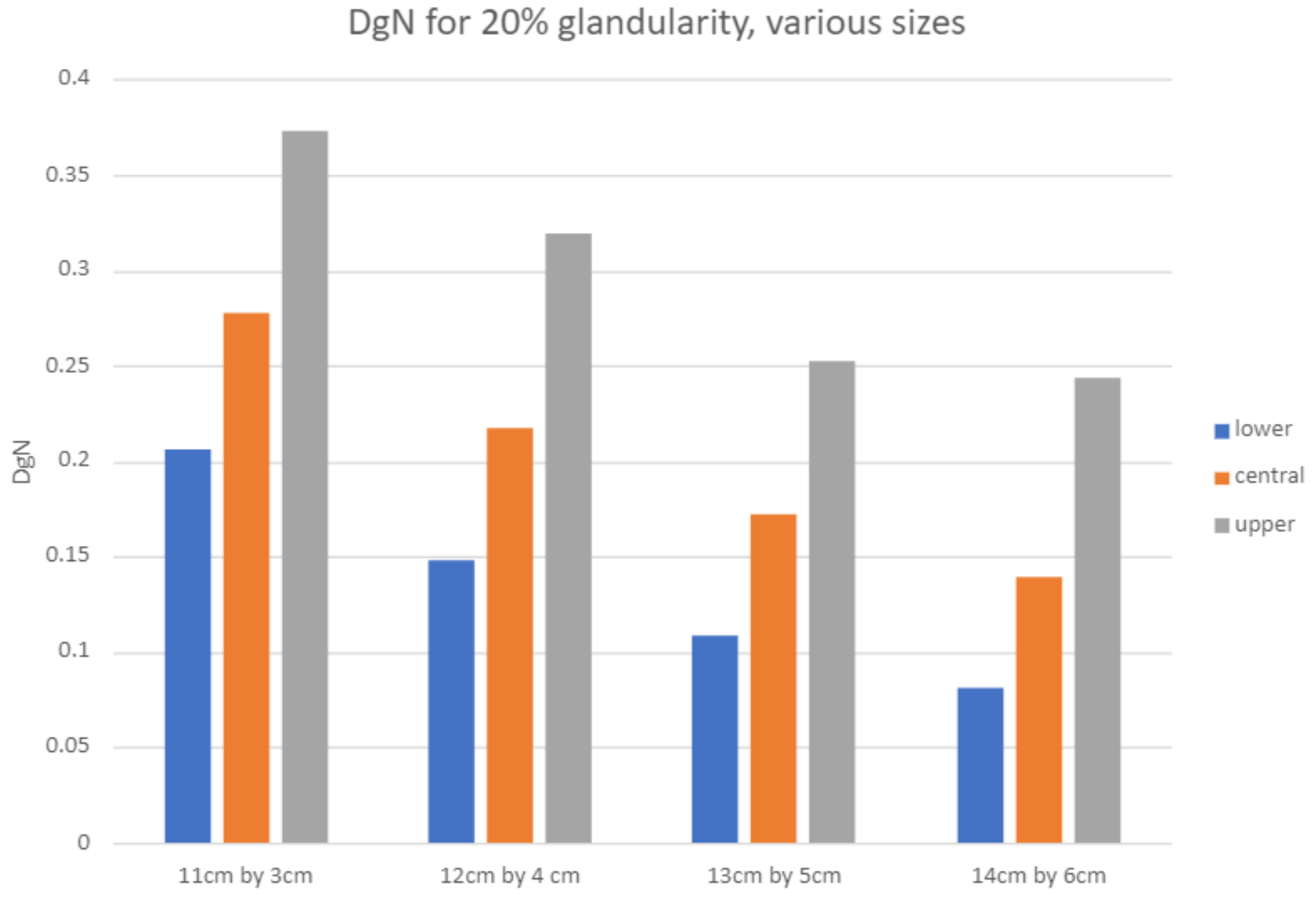

**Figure 6.** DgN values for 20% glandularity for breasts of various dimensions. Note that with increased breast thickness, the ratio between maximum and minimum DgN increases, with the ratio rising from 1.8 for a 3.29 cm thick breast to nearly 3.0 for a 6 cm

thick breast

Figure 7 shows DgN values as a function of glandular fraction for a fixed 4.29 cm thick breast. This thickness closely approximates the 4.2-cm standard or reference breast compressed thickness for the ACR accreditation phantom used for quality control and dosimetry [28-30]. It also lies within the clinically relevant compressed breast thickness range of 2–8 cm reported based on clinical datasets [16].

While the DgN remains approximately constant despite increasing glandularity, for the case where the fibroglandular tissue is concentrated at the center of the breast, the range between minimum and maximum DgN narrows. This is expectedly a reduced sensitivity to fibroglandular depth distribution when glandular tissue occupies a larger fraction of total breast volume.

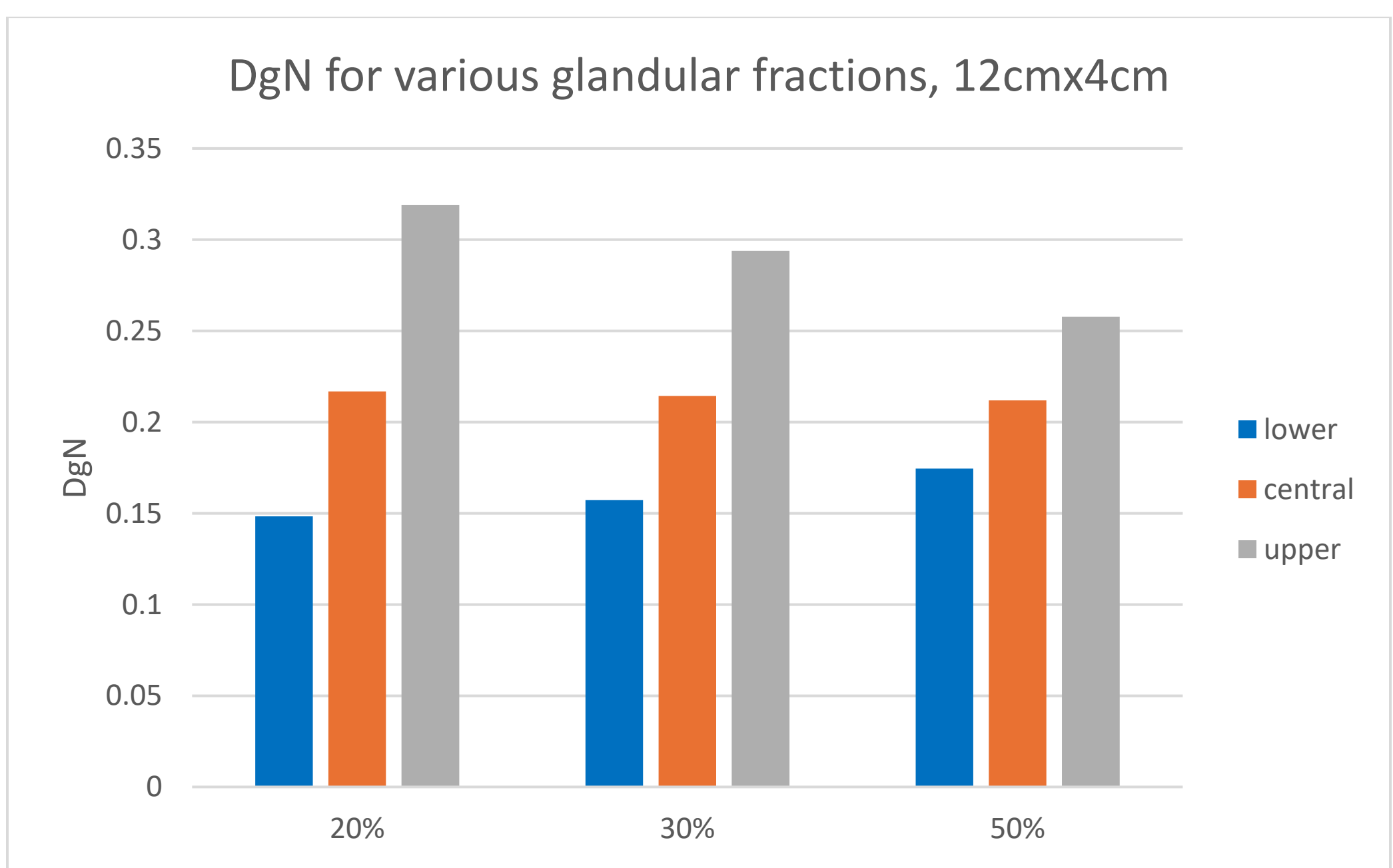

**Figure 7.** DgN values for varying glandular fractions, breast dimension including skin 12.29cm x 4.29cm. The difference between maximum and minimum is reduced as the glandular fraction is increased and occupies a larger fraction of the breast volume.

### 3.2 Results of Dosimetry from Breast Volume Reconstructions from Glandular fraction data

After reconstructing the breast object using ray tracing to place the fibroglandular tissue in the top (maximize DgN), bottom (minimize DgN) and center (likely DgN), consistent with the glandular fraction map, the voxelized phantoms are placed in TOPAS to ascertain the values of DgN after reconstruction and to compare with our original values with smooth TOPAS-generated objects. The TOPAS geometry is maintained as described in Section 2.1, with the breast object inserted as the voxelized reconstructed phantom (composed of thousands of 1 $mm^3$ voxels, each assigned as adipose tissue, fibroglandular tissue, skin, or air, as determined by the ray-tracing algorithm).

The reconstructed simulated breast object DgN values are compared against the original simulated breast objects (smooth analytical versions used in TOPAS). This evaluates the errors due to voxelizations and other digitization errors in the reconstruction process. The numerical DgN results are tabulated in Appendix A2 for 4 different sizes. The bolded results (where original and reconstruction are in the same configuration, that is either lower, central or upper) are shown in Figure 8, for comparison. For all 18 baseline cases, the true DgN from the original breast configuration (centrally located) lay within the reconstructed bounds. The largest deviation between the expected DgN and the reconstructed DgN occurs for the 13 cm × 5 cm case when the fibroglandular tissue is concentrated at the top of the breast near the source; in this instance, the discrepancy is about 8%.

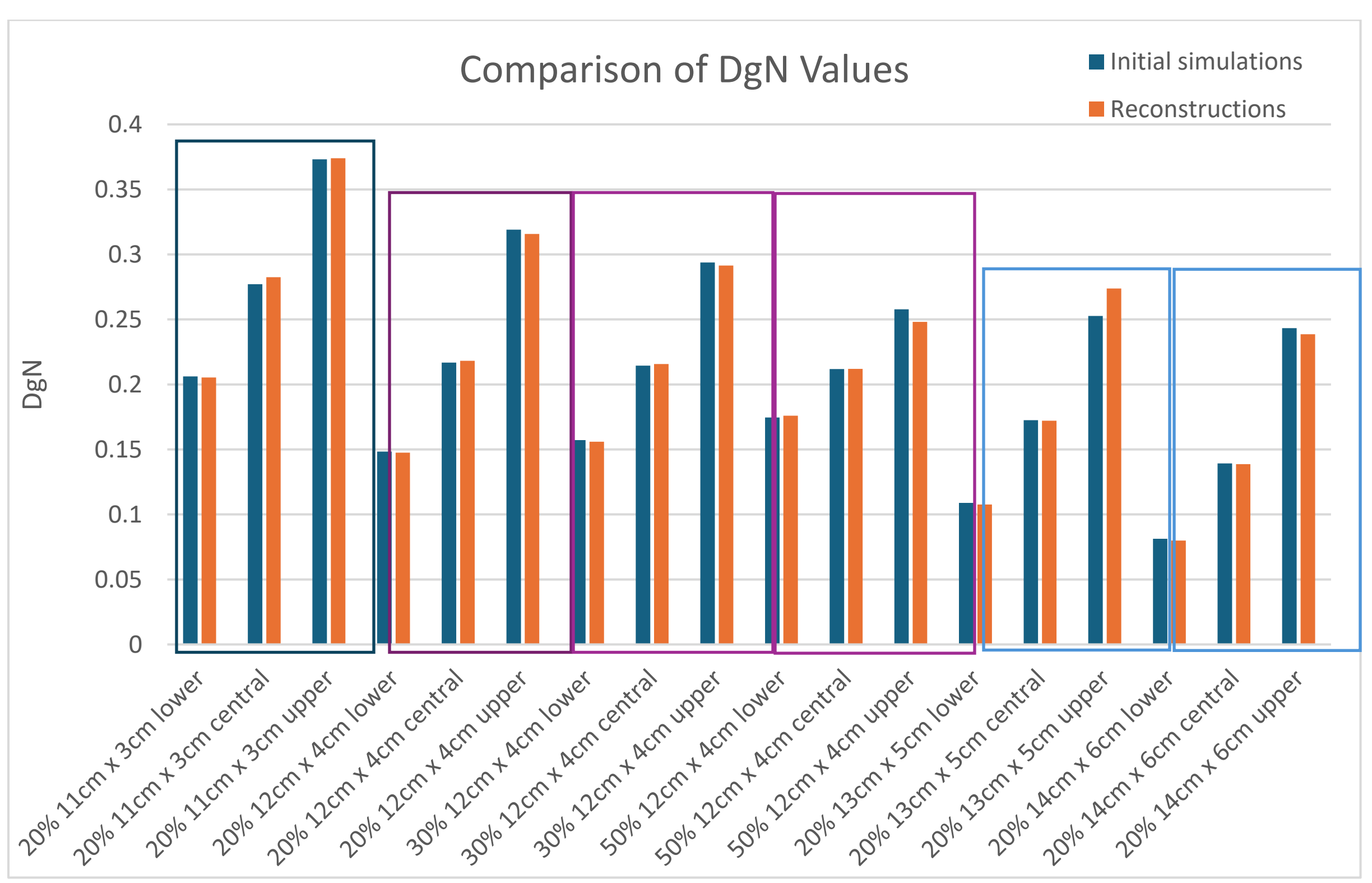

**Figure 8.** Comparing DgN values from the initial (original) simulations and the corresponding reconstruction. Groups of three (lower, central and upper) are grouped with rectangular boxes (also color-coded for size) for readability. The "initial" or orignal glandular DgN are from using smooth TOPAS generated objects while the "reconstructions" are the DgN from reconstructed voxelized phantoms, using the glandular fractions. The differences arise mainly from skin-layer approximations and ray voxelizations.

### 3.3 Estimates of Dosimetry errors due to Glandular fraction errors

Table 2 shows the analytically calculated dosimetry ratios when the glandular fraction is estimated to be ±5% different from its true value.

**Table 2. Dose Errors for +/-5% Glandular Fraction Errors**

| Nominal GF | Thickness (mm) | LowerGF/Nominal | HigherGF/Nominal | %ErrLower | %ErrHigher |
|---|---|---|---|---|---|
| 20% | 30 | 0.9554 | 1.0513 | -4.46% | 5.13% |
| | 40 | 0.9522 | 1.0477 | -4.78% | 4.77% |
| | 50 | 0.9550 | 1.0511 | -4.50% | 5.11% |
| | 60 | 0.9519 | 1.0482 | -4.81% | 4.82% |
| | | | | | |
| 30% | 30 | 0.9501 | 1.0567 | -4.99% | 5.67% |
| | 40 | 0.9517 | 1.0484 | -4.83% | 4.84% |
| | 50 | 0.9572 | 1.0526 | -4.28% | 5.26% |
| | 60 | 0.9529 | 1.0536 | -4.71% | 5.36% |
| | | | | | |
| 50% | 30 | 0.9570 | 1.0534 | -4.30% | 5.34% |
| | 40 | 0.9480 | 1.0529 | -5.20% | 5.29% |
| | 50 | 0.9472 | 1.0621 | -5.28% | 6.21% |
| | 60 | 0.9446 | 1.0637 | -5.54% | 6.37% |

The resulting dose error is approximately within 4.4–6.4%. This reflects the dependance of the dose estimate on the amount of glandular tissue present: since dose is accumulated only within glandular tissue, the dose error is expected to scale approximately with errors in glandular fraction estimation.

### 3.4 Dose Ratios for random distributions (MLO and CC-views) with respect to central placement

Table 3 summarizes the mean and standard deviation of the dose ratios obtained from 100 random draws sampled from the MLO and CC glandular distributions shown in Figure 4(a), relative to the central-placement estimate. Figures 9 and 10 illustrate the dependence of these ratios on breast thickness. For the vast majority of cases, the mean MLO dose ratio is less than unity and remains within 5% of the central-placement estimate. For a 60-mm breast with a 30% glandular fraction, the ratio is nearly unity, with an average difference of less than 0.5%. These results indicate that the central-placement approach provides an accurate yet conservative (slightly overestimated) approximation of the average true dose for the MLO case. For the CC case, the central-placement estimate can deviate by as much as ~14%. However, the dose ratio remains below unity in all cases, indicating that the central-placement method still provides a safe approximation.

**Table 3. Dose Ratio for MLO and CC Random Distributed versus Central Placement**

| GF (%) | Thickness (mm) | MLO Mean | MLO Std | CC Mean | CC Std |
|---|---|---|---|---|---|
| 20% | 30 | 0.9535 | 0.040 | 0.9002 | 0.038 |
| | 40 | 0.9631 | 0.046 | 0.8956 | 0.042 |
| | 50 | 0.9629 | 0.050 | 0.8814 | 0.044 |
| | 60 | 1.0039 | 0.063 | 0.9057 | 0.055 |
| | | | | | |
| 30% | 30 | 0.9536 | 0.032 | 0.8979 | 0.030 |
| | 40 | 0.9648 | 0.036 | 0.8929 | 0.033 |
| | 50 | 0.9744 | 0.043 | 0.8874 | 0.038 |
| | 60 | 1.0040 | 0.047 | 0.8997 | 0.041 |
| | | | | | |
| 50% | 30 | 0.9600 | 0.025 | 0.8948 | 0.024 |
| | 40 | 0.9617 | 0.028 | 0.8786 | 0.026 |
| | 50 | 0.9595 | 0.032 | 0.8595 | 0.028 |

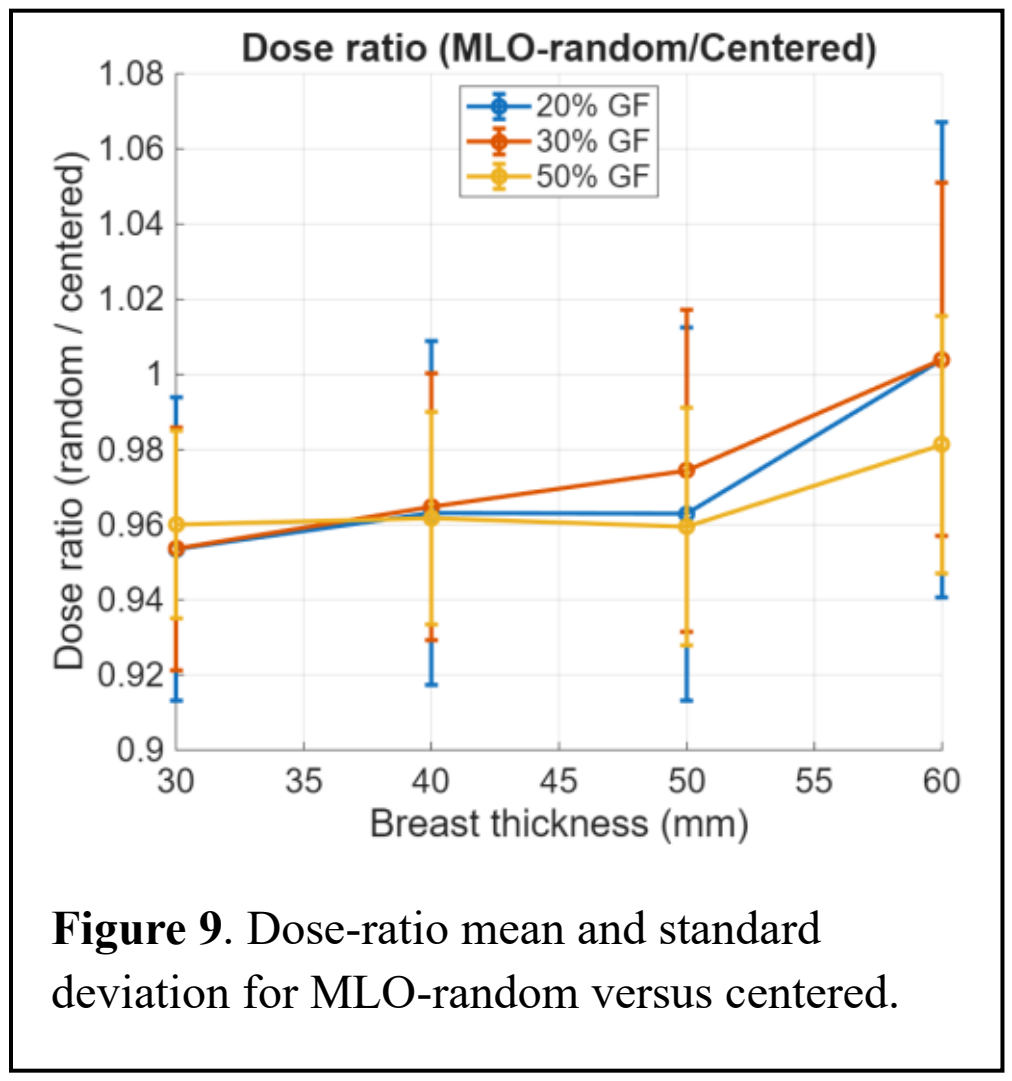

**Figure 9**. Dose-ratio mean and standard deviation for MLO-random versus centered.

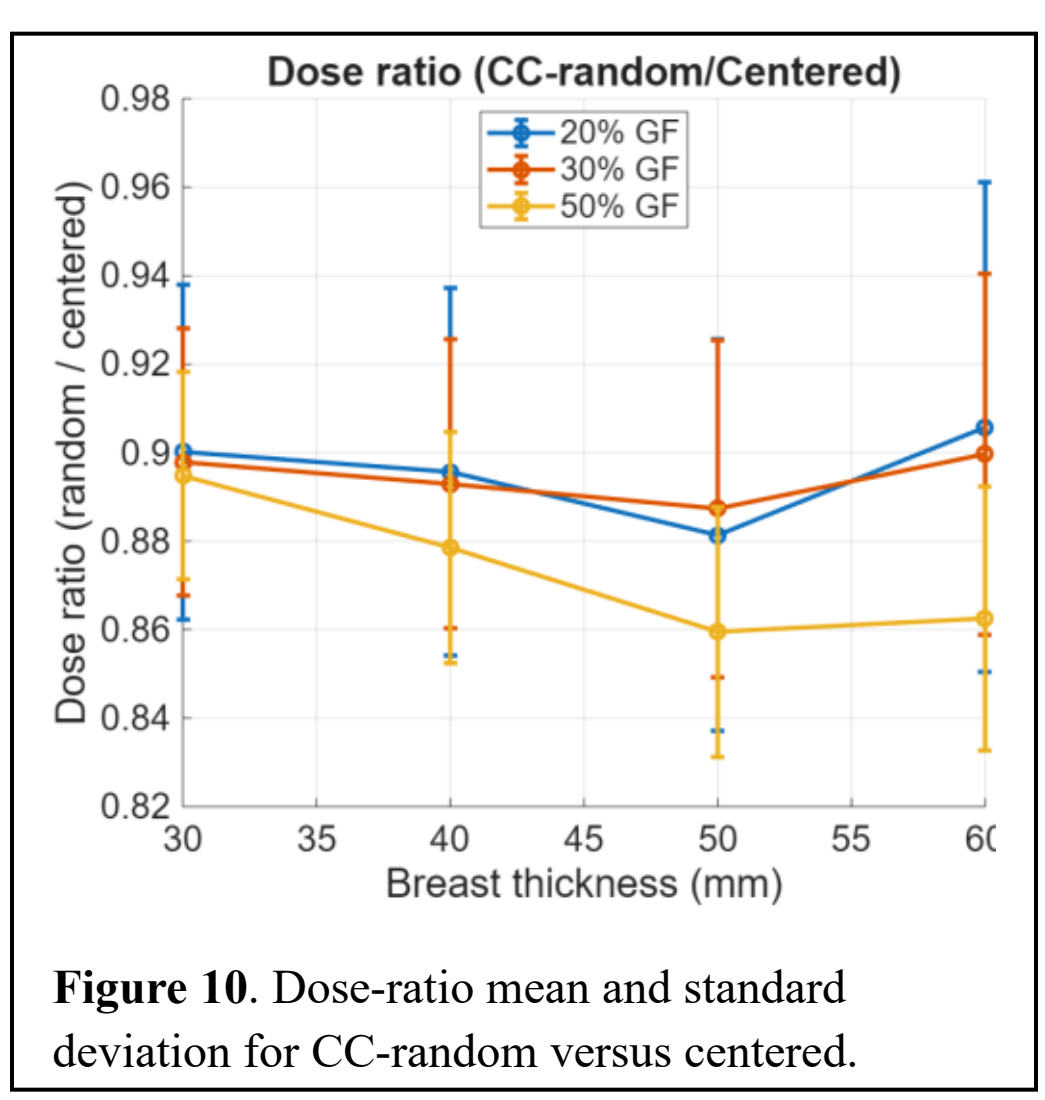

**Figure 10**. Dose-ratio mean and standard deviation for CC-random versus centered.

Table 4 shows the results obtained when the glandular tissue is placed at the centroid of each distribution (listed in the corresponding centroid column). In contrast to the central-placement case, centroid placement underestimates the dose for both the MLO and CC distributions, with dose ratios consistently greater than unity and deviations ranging from 8% to 25%. This behavior is consistent with the underlying glandular tissue distributions shown in Figure 4(a). While most of the glandular tissue is concentrated toward the caudal end, driving the centroid deeper into the breast, the distributions also exhibit a significant "front" tail in the shallower region. As a result, the randomly sampled placements are, on average, shallower than the centroid, leading to a higher average dose than the centroid-based estimate.

**Table 4. Dose Ratio for MLO and CC Random Distributed versus Centroid Placement**

| GF (%) | Thickness (mm) | MLO Centroid (mm) | MLO Mean | MLO Std | CC Centroid (mm) | CC Mean | CC Std |
|---|---|---|---|---|---|---|---|
| 20% | 30 | 17.0 | 1.0838 | 0.046 | 17.8 | 1.0791 | 0.045 |
| | 40 | 22.6 | 1.1372 | 0.054 | 23.7 | 1.1333 | 0.053 |
| | 50 | 28.3 | 1.1709 | 0.060 | 29.7 | 1.1657 | 0.059 |
| | 60 | 33.9 | 1.2627 | 0.080 | 35.6 | 1.2494 | 0.076 |
| | | | | | | | |
| 30% | 30 | 17.0 | 1.0838 | 0.037 | 17.8 | 1.0762 | 0.036 |
| | 40 | 22.6 | 1.1393 | 0.042 | 23.7 | 1.1301 | 0.041 |
| | 50 | 28.3 | 1.1854 | 0.052 | 29.7 | 1.1745 | 0.050 |
| | 60 | 33.9 | 1.2640 | 0.059 | 35.6 | 1.2427 | 0.056 |
| | | | | | | | |
| 50% | 30 | 17.0 | 1.0918 | 0.028 | 17.8 | 1.0735 | 0.028 |
| | 40 | 22.6 | 1.1380 | 0.033 | 23.7 | 1.1149 | 0.033 |
| | 50 | 28.3 | 1.1714 | 0.039 | 29.7 | 1.1432 | 0.038 |
| | 60 | 33.9 | 1.2421 | 0.043 | 35.6 | 1.2002 | 0.042 |

## 4. DISCUSSION

This work demonstrates that substantial uncertainty (upto a three-fold variation) in normalized glandular dose can arise solely from unknown depth-wise distribution of fibroglandular tissue, even when the breast thickness, glandularity, and projection appearance remain fixed or similar. The observed variation in DgN, particularly for thicker breasts, highlights the limitations of conventional homogeneous-50/50-breast dose models and emphasizes the importance of patient-specific dose characterization.

The proposed reconstruction-based framework (Section 3.2) was verified to provide accurate DgN values (maximum deviation 8% with respect to smooth objects in Section 3.1) and provides a practical method for estimating realistic bounds on DgN from a single mammographic projection image. Rather than attempting to assume a three-dimensional geometry, the method acknowledges inherent uncertainty in mammography and quantifies its dosimetry implications. This range-based approach is particularly relevant for thicker and less glandular breasts, where dose variability is greatest.

The differences (Figure 8) between the original and reconstructed volumes arise primarily from boundary voxelization and minor discrepancies in the skin layer. Overall, the results are similar, except for the upper-position cases, where boundary-voxelization effects and skin-layer differences are more pronounced. These effects can change the amount of glandular tissue located near the entrance side of the breast, where dose deposition is highest, leading to the observed differences.

The upper-and-lower bounds represent unlikely distributions, but show the possible bounds of dose for the particular patient.

Our analytical calculations based on random draws from TG-282 recommended distributions suggest that central placement may be an accurate (within 5% on average) and conservative estimate (overestimate) of the dose on average for the MLO case. For the CC-case the error was higher (within 15% on average) but the center-estimate remained an overestimate.

The centroid placement on the other hand, had 8-25% error and underestimated the dose on average compared to randomly drawn trials and may not represent a safe choice. The consistent underestimation could be explained by the significant "front tail" of the distribution, which contributes to a higher average dose than that predicted by the centroid-based estimate.

A major limitation of the analytical calculations is that Compton scatter is not considered. However, this limitation is somewhat mitigated by the fact that scatter contributes to both the numerator and denominator when dose ratios are calculated. The bulk of the scatter is expected to be a function of the water content of the tissue. Although it will also depend on the glandular distribution, the overall effect on the ratios is expected to be smaller.

Another limitation of the analytical calculations arises from the adaptation of the glandular tissue distributions recommended by TG-282 [17]. The Fedon et al. densities [18] used in that work were derived for a maximum breast depth of 70 mm. In this work, to extend the analysis to breasts of different compression thicknesses, the distributions were scaled to the corresponding breast depth while preserving their relative shape. This assumes that the normalized depth-wise distribution of glandular tissue remains unchanged with breast thickness. In practice, however, the spatial distribution of fibroglandular tissue may vary with breast size and compression thickness, and therefore the scaled distributions may not fully represent the true glandular anatomy of all patients.

Idealized glandular fraction maps were used to decouple reconstruction performance from glandular fraction estimation error. In clinical practice, uncertainties in GF estimation may widen dose bounds. On the average the errors were +/-5% [19]. Our analytical calculations show that +/-5% average glandular fraction estimation error will yield about similar dosimetry errors (4-6% in our case).

This study assumed a W/Al target-filter combination, which is used by several mammography vendors, including Fujifilm and Siemens. However, other systems employ different spectra, such as the W/Ag spectrum used by Hologic. The X-ray spectrum can influence both glandular fraction estimation [19] and glandular dose calculations. In general, narrower spectra reduce beam-hardening artifacts and can improve the accuracy of glandular fraction estimation. For example, our previous work [19] demonstrated that the narrower W/Ag spectrum yielded more accurate glandular fraction estimates than the W/Al spectrum.

One of the reasons for systematic glandular fraction errors is the beam-hardening effect of the polychromatic spectrum [19]. Our ongoing work involves improving the glandular fraction estimation model in [19] to take into account the full polychromatic spectrum rather than adopting the effective attenuation approximation used in [19]. The impact of different target-filter combinations on both glandular fraction estimation and patient-specific dose assessment will also be studied. Future work will focus on extending the framework to tomosynthesis acquisitions.

## 5. CONCLUSION

We have presented a method for estimating feasible ranges of normalized glandular dose in mammography using a pixel-wise glandular fraction map. This map can be derived from the projection image as described in our prior work. Monte Carlo simulations demonstrate that DgN may have high variation due solely to fibroglandular tissue depth distribution. The proposed reconstruction-based approach provides a practical framework for estimating bounds on DgN for mammographic images. Based on random trials from realistic distributions, analytical calculations of dose show that a central-placement of fibroglandular tissue is a conservative estimate (overestimate) of the dose, to within 5% for MLO and 15% for CC on average. The methods described represents a step toward more personalized and accurate dosimetry in breast cancer screening.

**ACKNOWLEDGEMENT**

The work presented here is based on the first author Lacey Medlock's MSc thesis, conducted under the supervision of Joyoni Dey [25]. This work was supported in part by NIH NIBIB Trail-blazer Award 1-R21-EB029026-01

**Author contributions:** **LLM** performed the TOPAS Monte-Carlo studies, and conducted data collection and analysis. **MST** contributed to analytical dosimetry computations. **BS** contributed to the TOPAS simulations. **JD** conceived the ideas of the study, contributed to the mathematical proofs, analytical dosimetry computations, and provided overall supervision. All authors reviewed and approved the final manuscript.

**Funding:** This work was supported in part by NIH NIBIB Trail-blazer Award 1-R21-EB029026-01 A1.

**Data Availability:** The data generated and analyzed during this study are available from the corresponding author upon reasonable request.

**APPENDIX A.1 : Proof of dose minimization/maximization localization**

We wish to prove that the dose for the case where the fibroglandular tissue is wholly concentrated near the top of the breast is greater than or equal to the dose for the case where the fibroglandular tissue is split into two segments: one at the top of the breast and one separated from the first by an area of adipose tissue. The net length and the fibroglandular length remain constant.

Dose is given as:

$$\dot{D} = \frac{kSE\frac{\mu_t}{\rho}e^{-\mu l}}{4\pi r^2} \qquad A.1$$

k, S, E, and 4π are constant. $\mu_t/\rho$ is dependent on the target material, which, in this case is always fibroglandular tissue, so it is constant. The dose then, is proportional to the remaining variables.

$$\dot{D} \propto \frac{e^{-\mu l}}{r^2} \qquad A.2$$

$\dot{D}$ is the dose rate. We can substitute this for dose, since the photons are delivered at a constant rate.

$$D \propto \frac{e^{-\mu l}}{r^2} \qquad A.3$$

r is the distance from the source to the dose point, which is the distance from the source to the surface, s, added to the distance from the surface to the point of interest, l.

$$D \propto \frac{e^{-\mu l}}{(l+s)^2} \quad \text{A.4}$$

Let $\mu_f$ and $\mu_a$ indicate the fibroglandular and adipose attenuation coefficients. The shield can be adipose tissue, fibroglandular tissue, or both, the equation has to be re-written specific to the voxel. Figure A1(a) and (b) shows two scenarios. In Figure A.1(a) there is fibroglandular tissue all clumped at the top (left in figure), near the source. In Figure A.1(b) there is some adipose tissue in between.

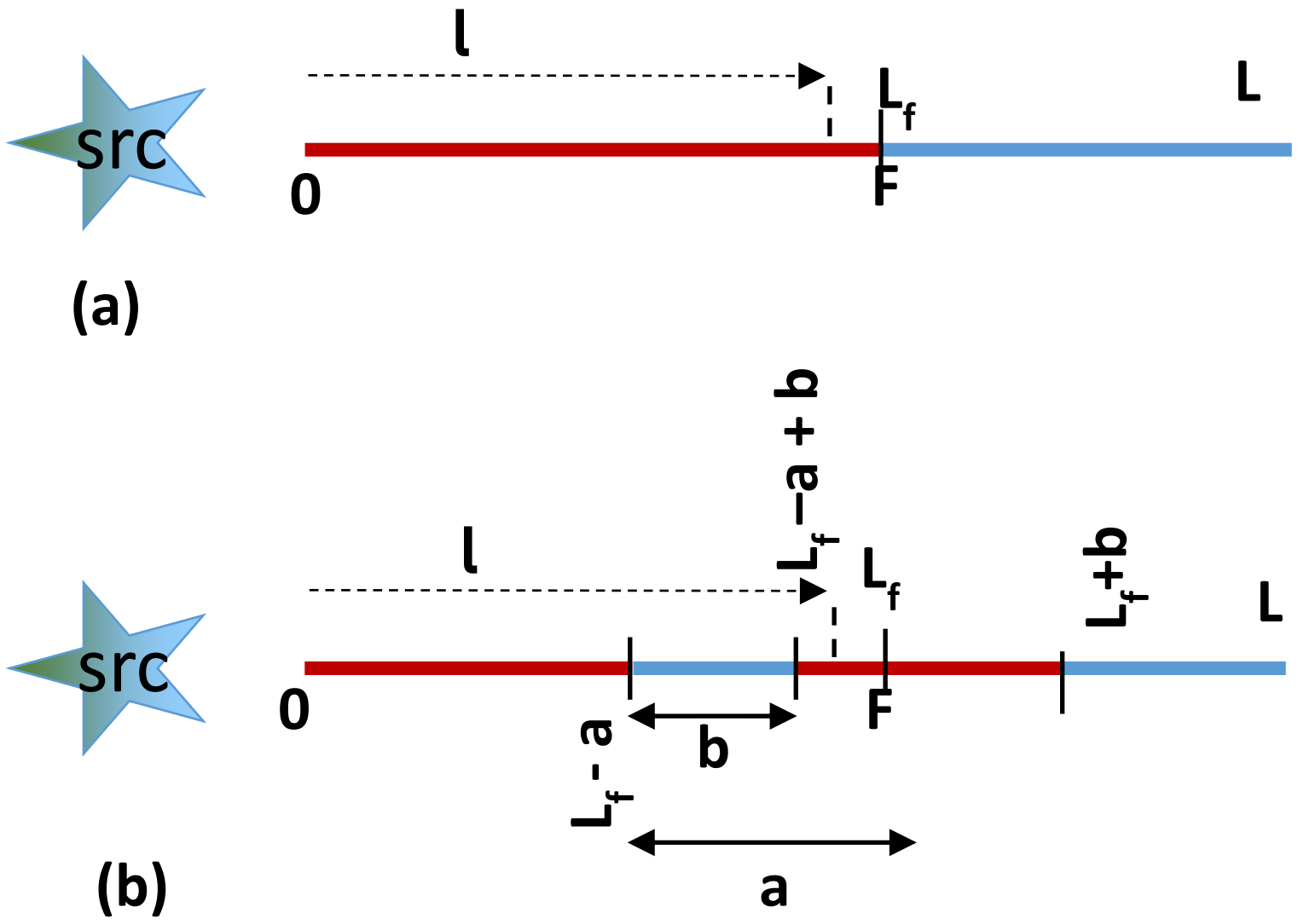

Figure A.1. (a) Fibroglandular (red) is at the top, near the source. The length variable ls small "l". (b) Fibroglandular tissue is fragmented with a layer of adipose of length b in between. The length variable is small "l".

When the fibroglandular tissue is all concentrated at the top of the breast, as depicted in Figure A.1(a), we can calculate the dose given to any one voxel at length *l* from top (left) as

$$D_1 \propto \frac{e^{-\mu_f l}}{(l+s)^2}$$

Then across all voxels

$$D_1 \propto \int_0^{L_f} \frac{e^{-\mu_f l}}{(l+s)^2} dl$$

Where $L_f$ is the length of fibroglandular tissue in Figure A1(a).

When the fibroglandular tissue is split into two segments, as depicted in Figure A.1(B), one at the top of the breast and one separated from the first by a layer of adipose tissue of length "b", the dose is proportional to

$$D_2 \propto \int_0^{L_f-a} \frac{e^{-\mu_f l}}{(l+s)^2} dl + \int_{L_f-a+b}^{L_f+b} \frac{e^{-[\mu_f(L_f-a)+\mu_a b+\mu_f(l-(L_f-a+b))]}}{(l+s)^2} dl$$

Where a is a number between 0 and $L_f$, and b is a number greater than 0.

This simplifies to

$$D_2 \propto \int_0^{L_f-a} \frac{e^{-\mu_f l}}{(l+s)^2} dl + e^{-\mu_a b} \int_{L_f-a+b}^{L_f+b} \frac{e^{-\mu_f (l-b)}}{(l+s)^2} dl$$

Note we deliberately kept the $e^{\mu_f b}$ term inside the integral. Changing variable, $l' = l - b$, we get

$$D_2 \propto \int_0^{L_f-a} \frac{e^{-\mu_f l}}{(l+s)^2} dl + e^{-\mu_a b} \int_{L_f-a}^{L_f} \frac{e^{-\mu_f l'}}{(l'+b+s)^2} dl'$$

Since $e^{-\mu_a b} \leq 1$ and $1/(l'+s+b) \leq 1/(l'+s)$ for $b \geq 0$, we will get

$$D_2 \leq \int_0^{L_f-a} \frac{e^{-\mu_f l}}{(l+s)^2} dl + \int_{L_f-a}^{L_f} \frac{e^{-\mu_f l'}}{(l'+s)^2} dl'$$

We see RHS = $\int_0^{L_f-a} \frac{e^{-\mu_f l}}{(l+s)^2} dl + \int_{L_f-a}^{L_f} \frac{e^{-\mu_f l}}{(l+s)^2} dl = \int_0^{L_f} \frac{e^{-\mu_f l}}{(l+s)^2} dl = D_1$

Therefore, $\boldsymbol{D_2 \leq D_1}$.

***Corollary 1:*** If there are multiple adipose segments for the same net length of fibroglandular tissue, we can apply this process multiple times across appropriate sub-lengths and we will obtain the same results.

***Corollary 2:*** By very similar mathematics or logic we can prove that when the fibroglandular tissue is concentrated at the bottom (right in our diagram), the dose to the fibroglandular tissue will be lower than any other configuration for the given net length of fibroglandular and adipose tissue.

## APPENDIX 2.2: Table of DgN from Reconstructed volumes

**Table A.2:** Comparison of actual DgN values for simulated breasts from Monte Carlo simulations with DgN values obtained by reconstruction. Original simulated objects are shaded and their DgN values bolded. Reconstructed objects are not shaded, and the bolded reconstruction DgN values are those which matched the fibroglandular location of the original object. The DgN calculations for 20% is shown for 4 breast sizes. The GF 30-50% is shown for one size. For readability the 4 sizes are separated into sub-tables. ***The bolded values are compared separately in Figure 8 in main text.***

| Object Size and glandular fraction | Location of fibroglandular tissue (original object) | Dose minimized (lower), central, or maximized (upper) for corresponding reconstructed object | DgN |
|---|---|---|---|
| 11cm x 3cm, 20% | Upper | | **0.3731** |
| | | Minimized | 0.2054 |
| | | Central | 0.2823 |
| | | Maximized | **0.3740** |
| 11cm x 3cm, 20% | Central | | **0.2771** |
| | | Minimized | 0.2055 |
| | | Central | **0.2825** |
| | | Maximized | 0.3744 |
| 11cm x 3cm, 20% | Lower | | **0.2062** |
| | | Minimized | **0.2053** |
| | | Central | 0.2824 |
| | | Maximized | 0.3746 |

| Object Size and glandular fraction | Location of fibroglandular tissue (original object) | Dose minimized (lower), central, or maximized | DgN |
|---|---|---|---|

| | | (upper) for corresponding reconstructed object | |
|---|---|---|---|
| 12cm x 4cm, 20% | Upper | | **0.3190** |
| | | Minimized | 0.1473 |
| | | Central | 0.2179 |
| | | Maximized | **0.3157** |
| 12cm x 4cm, 20% | Central | | **0.2168** |
| | | Minimized | 0.1475 |
| | | Central | **0.2181** |
| | | Maximized | 0.3161 |
| 12cm x 4cm, 20% | Lower | | **0.1484** |
| | | Minimized | **0.1475** |
| | | Central | 0.2184 |
| | | Maximized | 0.3166 |
| 12cm x 4cm, 30% | Upper | | **0.2938** |
| | | Minimized | 0.1557 |
| | | Central | 0.2155 |
| | | Maximized | **0.2914** |
| 12cm x 4cm, 30% | Central | | **0.2144** |
| | | Minimized | 0.1559 |
| | | Central | **0.2157** |
| | | Maximized | 0.2917 |
| 12cm x 4cm, 30% | Lower | | **0.1572** |
| | | Minimized | **0.1559** |
| | | Central | 0.2157 |
| | | Maximized | 0.2917 |
| 12cm x 4cm, 50% | Upper | | **0.2577** |
| | | Minimized | 0.1756 |
| | | Central | 0.2118 |
| | | Maximized | **0.2482** |
| 12cm x 4cm, 50% | Central | | **0.2119** |
| | | Minimized | 0.1758 |
| | | Central | **0.2119** |
| | | Maximized | 0.2483 |
| 12cm x 4cm, 50% | Lower | | **0.1746** |
| | | Minimized | **0.1759** |
| | | Central | 0.2121 |
| | | Maximized | 0.2485 |

| **Object Size and glandular fraction** | **Location of fibroglandular tissue (original object)** | **Dose minimized (lower), central, or maximized (upper) for corresponding reconstructed object** | **DgN** |
|---|---|---|---|
| 13cm x 5cm, 20% | Upper | | **0.2527** |
| | | Minimized | 0.1072 |
| | | Central | 0.1717 |
| | | Maximized | **0.2738** |
| 13cm x 5cm, 20% | Central | | **0.1724** |
| | | Minimized | 0.1075 |
| | | Central | **0.1721** |
| | | Maximized | 0.2742 |
| 13cm x 5cm, 20% | Lower | | **0.1088** |
| | | Minimized | **0.1076** |
| | | Central | 0.1720 |
| | | Maximized | 0.2743 |

| **Object Size and glandular fraction** | **Location of fibroglandular tissue (original object)** | **Dose minimized (lower), central, or maximized** | **DgN** |
|---|---|---|---|

| | | **(upper) for corresponding reconstructed object** | |
|---|---|---|---|
| 14cm x 6cm, 20% | Upper | | **0.2433** |
| | | Minimized | 0.0796 |
| | | Central | 0.1383 |
| | | Maximized | **0.2386** |
| 14cm x 6cm, 20% | Central | | **0.1392** |
| | | Minimized | 0.0798 |
| | | Central | **0.1387** |
| | | Maximized | 0.2390 |
| 14cm x 6cm, 20% | Lower | | **0.0812** |
| | | Minimized | **0.0800** |
| | | Central | 0.1389 |
| | | Maximized | 0.2394 |